%% file: ms.tex
\newif\ifastroph
\astrophfalse
\astrophtrue

\newif\ifsuppl
\supplfalse

\newif\ifcheck
\checkfalse

\ifastroph
  \documentclass{natureastroph}
\else
  \ifsuppl
     \documentclass{naturesup}
  \else
     \documentclass{nature}
       
  \fi
\fi
\def\S4figure{\ifastroph\else S\fi}

\newcommand\fignature[4]{
  \begin{figure}
    \ifastroph
    \centering \includegraphics[scale=#2]{#1}
    \caption{#3\label{#4}}
    \else
    \caption{
      \ifsuppl
      \includegraphics[scale=#2]{#1}
      \fi
      #3\label{#4}
    }
    \fi
  \end{figure}
}

\newcommand\ion[2]{#1$\;${%
\ifx\@currsize\normalsize\small \else
\ifx\@currsize\small\footnotesize \else
\ifx\@currsize\footnotesize\scriptsize \else
\ifx\@currsize\scriptsize\tiny \else
\ifx\@currsize\large\normalsize \else
\ifx\@currsize\Large\large
\fi\fi\fi\fi\fi\fi
\rmfamily#2}\relax}%

\pdfoutput=1
\usepackage{epstopdf}
\usepackage{amsmath}
\usepackage{rotating}

\usepackage{color}
\newcommand{\redpen}[1]{{\bf\textcolor{red}{#1}}}

\newcommand{\whitepen}[1]{{\bf\textcolor{white}{#1}}} 

\def\specMhi{EC1A}
\def\specMlo{EC1B}
\def\specDP{EC1C}

\def\reff@jnl#1{{\rm#1\/}}
\def\apj{\reff@jnl{Astrophys. J.}}
\def\apjl{\reff@jnl{Astrophys. J.}}               
\def\aj{\reff@jnl{Astron. J.}}                  
\def\araa{\reff@jnl{Annu. Rev. Astron. Astr.}}            
\def\apjs{\reff@jnl{Astrophys. J. Sup.}}              
\def\aap{\reff@jnl{Astron. Astrophys.}}               
\def\mnras{\reff@jnl{Mon. Not. R. Astron. Soc.}}      
\def\prd{\reff@jnl{Phys. Rev. D}}         
\def\prl{\reff@jnl{Phys. Rev. Lett.}}      
\def\pasp{\reff@jnl{Publ. Astron. Soc. Pac.}}              
\def\nat{\reff@jnl{Nature}}             

\def\arcdeg{\hbox{$^\circ$}}

\def\arcsec{\hbox{$^{\prime\prime}$}}

\def\snid{\ifmmode{\rm \tt SNID}\else{\tt SNID}\fi}
\def\dm15{\ifmmode{\Delta m_{15}}\else{$\Delta m_{15}$}\fi}

\def\magarcsec2{\ \rm{mag \ arcsec}^{-2}}
\def\ly{\ifmmode{\rm{ly}}\else{ly\fi}}

\title{Light echoes reveal an unexpectedly cool $\eta$~Carinae during its 19th-century Great Eruption}
 
\def\stsci{1}
\def\carnegie{2}
\def\princeton{3}
\def\hubble{4}
\def\stewart{5}
\def\lco{6}
\def\ucsb{7}
\def\cfa{8}
\def\mcmaster{9}
\def\jhu{10}
\def\clay{11}
\def\ctio{12}
\def\alma{13}
\def\catolica{14}
\def\london{15}

\author{A. Rest$^{\stsci}$, J.~L. Prieto$^{\carnegie,\princeton,\hubble}$,
N.~R. Walborn$^{\stsci}$, N. Smith$^{\stewart}$,
F.~B. Bianco$^{\lco,\ucsb}$, R. Chornock$^{\cfa}$,
D.~L. Welch$^{\mcmaster}$, D.~A. Howell$^{\lco,\ucsb}$,
M.~E. Huber$^{\jhu}$, R.~J. Foley$^{\cfa,\clay}$, W. Fong$^{\cfa}$,
B. Sinnott$^{\mcmaster}$, H.~E. Bond$^{\stsci}$,
R.~C. Smith$^{\ctio}$, I. Toledo$^{\alma}$, D. Minniti$^{\catolica}$, K. Mandel$^{\cfa,\london}$}

\begin{document}

\ifastroph
\else
  \ifsuppl
  \else
     \large
  \fi
\fi

\maketitle

\begin{affiliations}
\item Space Telescope Science Institute, 3700 San Martin Dr., Baltimore, MD 21218, USA
\item Carnegie Observatories, 813 Santa Barbara Street, Pasadena, CA 91101, USA
\item Department of Astrophysical Sciences, Princeton University, Peyton Hall, Princeton, NJ 08544
\item Hubble, Carnegie-Princeton  Fellow
\item Steward Observatory, University of Arizona, 933 North Cherry Avenue, Tucson, AZ 85721, USA
\item Las Cumbres Observatory Global Telescope Network, Goleta, CA 93117, USA
\item Department of Physics, University of California, Santa Barbara, CA 93106, USA
\item Harvard-Smithsonian Center for Astrophysics, 60 Garden Street, Cambridge, MA 02138, USA
\item Department of Physics and Astronomy, McMaster University, Hamilton,  \ifastroph \else Ontario, \fi L8S 4M1, Canada
\item Department of Physics and Astronomy, Johns Hopkins University, Baltimore, 3400 North Charles Street, MD 21218, USA
\item Clay Fellow
\item Cerro Tololo Inter-American Observatory, National Optical Astronomy Observatory, Colina el Pino S/N, La Serena, Chile
\item ALMA, KM 121 CH 23, San Pedro de Atacama, II Region, Chile 
\item Dept. of Astronomy and Astrophysics, Pontifica Universidad Catolica, 
Santiago 22, Chile
\item Imperial College London, Blackett Laboratory,
Prince Consort Rd, London SW7 2AZ, UK
\end{affiliations}

\input{main}
\ifastroph
  \else
  \ifsuppl
  \else

  \fi
\fi

\begin{addendum}

\item We thank R. Humphreys, K. Davidson, and J. Vink for comments and
discussions. We thank S. Blondin for help with the continuum
subtraction. 
The Blanco 4m telescope is a facility of the Cerro Tololo
Inter-American Observatory, National Optical Astronomy Observatory,
which are operated by the Association of Universities for Research in
Astronomy, under contract with the National Science Foundation.
We use data from the UVES Paranal Observatory Project.
The computations in this paper were run on the Odyssey cluster
supported by the FAS Science Division Research Computing Group at
Harvard University.  Observations were obtained at LCOGT, and FBB and
DAH acknowledge support from LCOGT.

\item[Author Contributions] All authors contributed to the drafting of
the paper.  A.R., N.S and R.C.S. imaged the area around $\eta$~Car.
A.R. and M.H. reduced the imaging data. H.E.B. provided images of the
echoes that guided our spectroscopic pointings.  J.L.P., R.C., R.J.F.,
and W.F. obtained the spectra and reduced them. A.R. and
J.P.L. performed spectral analysis and interpretation. A.R., N.R.W.,
and F.B.B performed spectral classification. F.B.B. and
K.M. correlated the spectra.  A.R., D.L.W. and B.S. modelled the light
echo. I.T. and D.M. provided imaging of $\eta$~Car. F.B.B and D.A.H
provided the FTS images, F.B.B and A.R. reduced them.

 \item[Author Information]  The authors declare that they have no
competing financial interests. Correspondence and requests for materials
should be addressed to A.R.~(arest@stsci.edu).

\end{addendum}

\ifastroph
  \input{si}

  \input{tables}
  \clearpage
  \bibliography{ms}
\else
  \ifsuppl
    \pagestyle{empty}
    \input{si}  
    \clearpage
    \input{tables}
    \clearpage
    \bibliography{ms}
    \clearpage
  \else
    \input{figures}  
  \fi
\fi

\end{document}

%% file: main.tex
{\bf 
$\eta$~Carinae is one of the most massive binary stars  in the
Milky Way\cite{DH97,Damineli96}. It became the second-brightest star
in the sky during its mid-19th century ``Great Eruption,'' but then
faded from view (with only
naked-eye estimates of brightness\cite{Frew04,SF11}).
Its eruption is unique among known astronomical transients in that it
exceeded the Eddington luminosity limit for 10 years.
Because it is only 2.3~kpc away, spatially resolved studies of the
nebula have constrained the ejected mass and velocity, indicating that
in its 19th century eruption, $\eta$~Car ejected more than
10~$M_{\odot}$ in an event that had 10\% of the energy of a typical
core-collapse supernova\cite{Smith03,Smith08} without destroying the
star.
%
Here we report the discovery of light echoes of $\eta$~Carinae which
appear to be from the 1838-1858 Great Eruption.  
Spectra of these light echoes show only absorption lines, which are
blueshifted by $-$210~km~s$^{-1}$, in good agreement with predicted
expansion speeds\cite{Smith08}.
The light-echo spectra correlate best with those of G2-G5 supergiant
spectra, which have effective temperatures of $\sim$5000~K.  In
contrast to the class of extragalactic outbursts assumed to be analogs
of eta Carinae's Great
Eruption\cite{Goodrich89,Humphreys94,Humphreys99,VanDyk00,Vink09,Kashi10_ngc300},
the effective temperature of its outburst is significantly cooler than
allowed by standard opaque wind models\cite{Davidson87}.  This
indicates that other physical mechanisms like an energetic blast wave
may have triggered and influenced the eruption.
}

\medskip

$\eta$~Car-like giant eruptions of Luminous Blue Variables (LBVs) are
characterized by significant mass-loss and an increase in luminosity
by several magnitudes\cite{Humphreys94,Humphreys99}.
Traditionally, it was thought
that this increase in luminosity 
drives a dense wind, producing an
optically-thick cooler pseudo-photosphere with a minimum effective
temperature of ~7000 K and an F-type spectrum\cite{Davidson87}.  Within
this model, $\eta$~Car has been considered the prototype of the
so-called ``supernova imposters''
\cite{Goodrich89,Humphreys94,Humphreys99,VanDyk00,Vink09,Kashi10_ngc300}.



We obtained images in proximity to $\eta$~Car
(Figure~\ref{fig:diffim}) that, when differenced, show a rich set of
light echoes. The largest interval between our images is 8 years.  We
have also found similar echo candidates at other positions, which we
are currently monitoring.  The large brightening and long duration
point to the Great Eruption as the source of the light echoes.  We
have also obtained a composite light curve in the SDSS $i$ filter of
the light echoes (see Fig.~\ref{fig:lc}),
showing a slow decline of several tenths of a magnitude over half a
year.  This light curve is most consistent with the historical
observations\cite{SF11} of a peak in 1843, part of the 1838-1858 Great
Eruption, although further observations are necessary to be certain
(see the Supplementary Information, SI).

Spectra of the light echoes (see Figure~\ref{fig:spec}) show
only absorption lines characteristic of cool stellar photospheres, but
no evidence of emission lines.  In 
particular the
\ion{Ca}{II} infrared (IR) triplet is only in absorption.
Because of bright ambient nebular emission, it is
difficult to determine if there is any H$\alpha$ emission from $\eta$~Car
itself, but in any case it must be weak if present.  By
cross-correlating each of our $\eta$~Car echo spectra with the UVES
spectral library\cite{Bagnulo03} (see
Figure~S2~and~S3) 
\ifcheck
  \redpen{CHECK S\ref{fig:speclines}~and~S\ref{fig:pdf}}
\fi
we find best agreement
with supergiant spectral types in the range of G2-G5, with an
effective temperature of $\sim$5000~K. Spectral types of F7 or earlier
are ruled out by our analysis (see SI for more details).

Doppler shifts of absorption features in the echo spectra provide direct
information about the outflow speeds during the eruption.
The \ion{Ca}{II} IR triplet absorption features in the spectrum are
noticeably blueshifted (see Figure~S2).
\ifcheck
  \redpen{CHECK S\ref{fig:speclines}}
\fi
By cross-correlation with G-type\cite{Cenarro01} templates, we determine
velocities of $-202\pm9$, $-210\pm14$, and $-237\pm17$~km~s$^{-1}$ for
our three individual spectra and an average
velocity of $-210\pm30$~km~s$^{-1}$, which includes an uncertainty for
the dust sheet velocity.

The bipolar nature of the Homunculus shows that the $\eta$~Car
Great Eruption was strongly aspherical.  It was previously predicted
that the outflow speeds one would derive from spectra of $\eta$~Car in
outburst, looking at the poles and equator of the double lobes, would
be $\sim-$650~km~s$^{-1}$ and $-$40-100~km~s$^{-1}$,
respectively\cite{Smith06} (outflow speeds near the equator have a
steep latitude dependence). The light echo we investigate in this
paper arises from latitudes near the equator of $\eta$~Car (see Figure~S1),
\ifcheck
  \redpen{CHECK S\ref{fig:3Dplot}}
\fi
and the measured blue-shifted velocity of $-210\pm30$~km~s$^{-1}$ is
in good agreement with expansion speeds within $\pm$20\arcdeg\ of the
equatorial plane.  We also find a strong asymmetry in the
\ion{Ca}{II} IR triplet, extending to a velocity of
$-$850~km~s$^{-1}$ - well below the speed of the fastest polar
ejecta found previously\cite{Smith08}, but is in good agreement with
speeds observed in the blast wave at lower latitudes\cite{Smith08}.
Future observations of light echoes viewing the $\eta$~Car eruption
from different directions, in particular from the poles, has the
potential to observe these very high-velocity ejecta and other
asymmetries.



A characteristic of LBV outbursts is their transition from a hot
quiescent state to a cooler outburst state, although this feature is
less well observed for the giant eruptions (see Figure~\ref{fig:hr}).
Two potential models for LBV outbursts involve either an opaque
stellar wind driven by an increase in luminosity, or a hydrodynamic
explosion. The traditional mechanism for $\eta$~Car-like 
giant
eruptions has been that an unexplained increase in luminosity drives a
denser wind, so that an optically thick pseudo-photosphere forms at a
layer much larger and cooler than the hydrostatic stellar
surface\cite{Davidson87}. This model predicts a minimum effective
temperature of 7000 K, resembling A or F-type
supergiants\cite{Humphreys94,Smith04}, since the wind opacity depends
on the temperature (see Figure~\ref{fig:hr}). They evidently occur as
a massive star attempts to evolve redward and encounter the
Humphreys-Davidson Limit\cite{Davidson87}, beyond which no stable
stars are observed.

Surprisingly, our G-type light-echo spectrum of the $\eta$~Car Great
Eruption is inconsistent with expectations of an opaque-wind
model\cite{Davidson87} (see Figure~\ref{fig:hr}). This model also
fails to explain the high 10$^{50}$ erg kinetic energy\cite{Smith03}
and the presence of a fast blast wave at large radii\cite{Smith08}.
Instead, these observations point toward a hydrodynamic explosion 
at least influencing the Great Eruption\cite{Damineli96,Smith03,Smith08}.

The first {\it visual} spectroscopic observations of $\eta$~Car around
1870 showed emission lines \cite{LeSueur70a,LeSueur70b}.  A
photographic spectrogram obtained during its Lesser Eruption circa
1890\cite{Walborn77,Humphreys08} resembles an F-type supergiant
blueshifted by $-$200~km~s$^{-1}$, with moderate hydrogen P~Cygni
profiles, which {\it is} as expected in the opaque-wind
model\cite{Davidson87}.  The difference between the 1890 and our
light-echo spectra of the Great Eruption is therefore quite striking,
indicating that two distinct physical processes may have been involved
for two outbursts of the same object.  However, the 1890 event also
produced a mass ejection, the Little Homunculus, with the same axial
symmetry as the Great Eruption\cite{Smith05}, albeit of a much smaller
amount.

LBV giant eruptions are rare, and only have been recorded in our
Galaxy in the last 400 years, the Great Eruption of $\eta$~Car and
P~Cygni's giant eruption in the 17th century.  Because of their
considerable intrinsic brightness just below the luminosity of faint
core-collapse SNe, about two dozen giant eruption candidates,
so-called SN imposters since they have been often mistaken for SNe,
have been found in various extragalactic SN
searches\cite{Goodrich89,Humphreys94,Humphreys99,VanDyk00,Vink09,Kashi10_ngc300}.
Typically, the hotter SN imposters have steep blue continua, stronger
and broader Balmer lines, and relatively weak absorption, whereas the
cooler ones tend to have redder continua, weaker and narrower Balmer
lines, strong [\ion{Ca}{II}] and \ion{Ca}{II} emission, deeper P~Cygni
absorption features, and in some cases stronger absorption spectra
similar to F-type supergiants\cite{Smith11}.  However, the $\eta$~Car
Great-Eruption light-echo spectrum is quite different.  Its spectral
type is G2-G5, significantly later than all other SN imposters at
peak.  Furthermore, the \ion{Ca}{II} IR triplet lines are only in
absorption.  
For the extreme mass-loss rates required in $\eta$~Car's Great
Eruption, another process must give rise to the apparent temperature.

$\eta$~Car's Great Eruption has been considered the prototype of the
extragalactic SN imposters or $\eta$~Car analogues, even though it is
actually an extreme case in terms of radiated energy (10$^{49.3}$
erg), kinetic energy ($>$$10^{50}$~erg), and its decade-long
duration\cite{Smith11}. The spectra of the light echo indicates now
that it is not only extreme, but a different and rather unique object.
It is difficult to see how strong emission lines could be avoided in
an opaque wind where the continuum photosphere is determined by a
change in opacity, and its temperature and broad absorption lines are
more consistent with the opaque cooling photosphere of an explosion.
The cause that triggered such an explosion and the mass-loss without
destroying the star is still unknown, but predictions from future
radiative transfer simulations trying to explain $\eta$~Car and its
Great Eruption can now be matched to these spectral observations.
Other alternative models that were proposed, e.g. the ones that use
mass accretion 
from the companion star during periapsis passage as a
trigger for the eruption\cite{Kashi10_ngc300}, can be either verified
or dismissed. 
\whitepen{\cite{Rest05b,Rest08b,Rest11_leprofile,Smith10,Foley11}}

%% file: si.tex
\clearpage
\renewcommand*{\thesection}{S\arabic{section}}

\section{Supplementary Information\label{sec:SI}}
\subsection{Three dimensional orientation}
The scattered-light path is shown in Figure~\S4figure\ref{fig:3Dplot}.  The
3D-plot shows that with this light echo we see the eruption of
$\eta$~Car from a viewing angle perpendicular to the principal axis of
the Homunculus Nebula. It will be very interesting to compare spectra
seen from this viewing angle with spectra from light echoes that view
the eruption from the poles of the bipolar outflow.

\subsection{Evidence that these are light echoes of the mid-19th century Great
Eruption of $\eta$~Car}

The angular separation between the center of our light-echo images and
$\eta$~Car is $\sim$0.5~degrees, as shown in the left panel of
Figure~1,
\ifcheck
  \redpen{CHECK \ref{fig:diffim}} 
\fi
in which our pointing is indicated with a white box. In the middle
panel of this figure, our three epochs of SDSS {\it i}-band imaging
obtained with the $\rm8k\times8k$ Mosaic imager on the CTIO 4-m Blanco
telescope are displayed. The images, each with an exposure of 160~s,
were obtained on 2003 Mar 10 ({\it A}), 2010 May 10 ({\it B}), and
2011 Feb 6 ({\it C}).  Difference images, $C-A$ and $C-B$, are also
shown (upper and middle right panels, respectively). In them, we
notice excess flux from the first (black) and second epochs (white),
which we interpret as light echoes from $\eta$~Car.  Sample light-echo
positions are indicated with blue (epoch {\it A}) and red (epoch {\it
B} and {\it C}) arrows.

Comparing the first- and second-epoch images, we can determine the
direction of the apparent motion of the light echo and infer the
direction to the illuminating source. Most of the excess flux in the
first-epoch image is toward the northern part of the image (blue
arrows in Figure~1,
\ifcheck
  \redpen{CHECK \ref{fig:diffim}})
\fi
whereas most of the excess flux in the second-epoch image is in the
southern portion (red arrows in Figure~1.
\ifcheck
  \redpen{CHECK \ref{fig:diffim}}) 
\fi
If the light echoes arise from $\eta$~Car, this is exactly the
directional sense expected. We have also found similar regions of
excess flux at other positions in proximity to $\eta$~Car, with
apparent motions consistent with $\eta$~Car being the source of the
outburst light, which we will discuss in a follow-up paper.  The
vector method we have used previously to determine the origin of the
light echoes for SNR 0509$-$67.5, Cas~A, and
Tycho\cite{Rest05b,Rest08b} was applied here and, not surprisingly, we
conclude that these excess fluxes are most likely light echoes from a
dramatic brightening of $\eta$~Car.

The flux profile of a light echo, which is the cut through the light
echo along the axis pointing toward the source event, is the projected
light curve, stretched or compressed depending on the inclination of
the scattering dust filament, and convolved with the effects of the
dust width and the seeing\cite{Rest11_leprofile}.  In pathological
cases where the inclination of the dust filament is very unfavorable
or where the dust filament is significantly thicker than any yet
observed, a useful correspondence may be
lost\cite{Rest11_leprofile}. For light echoes from sources in our
Galaxy scattered by dust with typical widths and inclinations, we
expect that the apparent motion of the light echo will move at a rate
of one projected echo width between two epochs separated by twice the
outburst event duration. For known galactic SNe light echoes the
timescale of the brightest phase is of the order of a few months, so
that an imaging separation time of 1 year is sufficient for there to
be no overlap (on the sky) of light-echo
features\cite{Rest05b,Rest08b}.  For a several-decade duration
outburst, like that of $\eta$~Car, the difference image $C-B$ has only
a time difference of 1 year, and can only reveal the small boundary
fractions of the light echo in $C$, whose more complete extent can be
seen in $C-A$ which has a time difference of 8 years.  We conclude
that the duration of the event illuminating these light echoes is
significantly longer than one year, consistent with our conclusion
that these are echoes of the 20-year long Great Eruption of
$\eta$~Car.

\subsection{Determination of spectral type}

We compare the light-echo spectra to the compilation of supergiant
spectra in the UVES atlas\cite{Bagnulo03} (see Table~S2).
\ifcheck
  \redpen{CHECK}
\fi
The upper left panel of Figure~\S4figure\ref{fig:speclines} shows the
light-echo spectra, flattened by dividing by low-order continuum fits,
and the UVES supergiant sequence.  The instrument configuration for
the EC1A spectrum was designed to optimize the S/N ratio in the
wavelength region of the Ca II IR triplet, resulting in a lower S/N
ratio for $\lambda<6000$~\AA. For this reason the EC1A spectrum is
absent from the upper left panel of Figure~\S4figure\ref{fig:speclines}.  The
light-echo spectra correlate very well with late-F and G-type stars,
in particular the Mg~b lines and line blends at 5270~\AA\ and
6497~\AA. Both earlier- and later-type spectra show significantly
fewer similarities.

We calculate the cross-correlation parameter $r$ between the light
echo and the UVES spectra using the IRAF routine {\tt xcsao} in the
wavelength range 5050-6500~\AA\ (see left panel of
Figure~\S4figure\ref{fig:pdf}). We exclude wavelength ranges contaminated by
fore/background emission lines
(see Table~S4). 
\ifcheck
  \redpen{CHECK} 
\fi
We find that extending the wavelength region to redder wavelengths
(e.g., 5050-7500~\AA) decreases the correlation, and in particular the
correlation differences between the different spectral types. The
reasons for a decrease in correlation are threefold: (1) the spectral
differences between F, G, and K stars are stronger in the bluer
wavelength ranges, and therefore including redder wavelengths dilutes
the discrimination power, (2) strong fore/background emission lines like
$H_{\alpha}$, \ion{N}{II}, and \ion{S}{II} introduce large
discontinuities in the wavelength coverage, and (3) the chip gaps fall
into the 6500-7500~\AA\ wavelength range, further adding to the
discontinuities. For the above reasons we determine the 5050-6500~\AA\
wavelength range to be optimal for the correlation analysis.



To quantitatively estimate the temperatures of the supergiants that
correlate best, we smooth $r(T_{eff})$ with a Gaussian of width
$\sigma=300$~K (see lines in left panel of Figure~\S4figure\ref{fig:pdf}). We
find that the temperature with maximum correlation is 5210~K and 4950K
for \specMlo\ and \specDP, respectively. The right panel of
Figure~\S4figure\ref{fig:pdf} shows the probability density functions (PDFs)
of the temperature, $T_{eff}$, of the supergiant spectral template
having the best smoothed correlation, $r(T_{eff})$, with each
light-echo spectrum. The PDFs are computed by bootstrap resampling of
the distribution $10^5$ times.  The 95\% confidence intervals of the
best-matching temperatures are 4850-5550~K (spectral type G0-G5) and
4450-5400~K (spectral type G0-K1) for \specMlo\ and \specDP,
respectively.  Using an extended wavelength range of 5050-7500~\AA,
the 95\% confidence intervals are 3700-6300~K (F8-K5) and 3450-5100~K
(G2-M1) for \specMlo\ and \specDP, respectively. For the reasons
mentioned above, the extended wavelength range results in a spectral
type that is less constrained compared to the 5050-6500~\AA\ range.
We conclude that the light-echo spectra agree best with supergiant
spectral types in the range of G2-G5,
\ifcheck
  \redpen{CHECK} 
\fi
with an effective temperature of $\sim$5000~K.
\ifcheck
  \redpen{CHECK} 
\fi
We conservatively exclude spectral types of F7 and earlier.


Unfortunately, the UVES spectra have a gap in coverage around the
\ion{Ca}{II} IR triplet, and therefore we use a compilation of
supergiant spectra\cite{Cenarro01} covering the \ion{Ca}{II} triplet
wavelength range (see Table~S3).
\ifcheck
  \redpen{CHECK}
\fi
The lower-left panel of Figure~\S4figure\ref{fig:speclines} compares the
observed light-echo spectrum with this atlas. For early-type stars up
to G0, the wavelength range redward of 8300~\AA\ is mainly dominated
by the H Paschen series, which contaminates the \ion{Ca}{II} triplet
but gradually disappears as the effective temperature
decreases\cite{Cenarro01}. Note that the observed light-echo spectra
do not show any clear signs of the H Paschen lines; in particular, the
lines at 8409~\AA\ and 8498~\AA\ are at most very weak.  The earliest
spectral types in agreement with such weak or non-existent H-Paschen
lines are late F-types, in excellent agreement with the spectral
correlation results in the wavelength range 5050-6500~\AA.

Visual observers noted a reddish or ``ruddy'' color during
$\eta$~Car's Great Eruption, e.g. described by Herschel as ``redder
than Arcturus''\cite{Frew04,SF11}. These observations point to a
temperature of $T_{eff}<4500$~K, lower than the temperature we infer
from the spectral lines. It is plausible that reddening by grains
forming in the eruption cause the apparent color. We note, however,
that the temperature inferred from the spectral type is much more
reliable than an apparent color, in particular one done visually,
since it is not influenced by unknown amounts of reddening caused by new
dust formation along the line of sight.



The lower-right panel of Figure~\S4figure\ref{fig:speclines} shows velocities
determined from cross-correlation from spectra of different spectral
types.  The contribution to the derived velocity caused by the motion of the
reflecting dust sheet, i.e. the \emph{moving-mirror effect}, is likely
less than 30~km~s$^{-1}$ given the relatively low dominant expansion
speeds of cool gas in the Carina Nebula H~{\sc ii}
region\cite{Walborn02}.  This systematic error could be constrained
better with more spectra of light echoes located close enough to each
other that they probe similar viewing angles, but different enough
that the scattering dust is parsecs apart and thus has independent
velocities.

\subsection{The most similar SN impostor}
The closest SN impostor in temperature is UGC~2773~OT2009-1, which is
also dominated by a forest of absorption lines, similar to an F-type
supergiant\cite{Smith10,Foley11}. However, the \ion{Ca}{II} IR triplet
has a P~Cygni profile with a strong emission component\cite{Foley11}.
Interestingly, UGC~2773~OT2009-1 is also one of the few examples where
the outburst has persisted for years and in fact still
continues\cite{Smith10,Smith11,Foley11}.

\subsection{H$\alpha$ line}
The interpretation of the H$\alpha$ emission line is difficult, since
the fore/background emission line subtraction is imperfect. The
left panel of Figure~3
\ifcheck
  \redpen{CHECK \ref{fig:spec}} 
\fi
shows the H$\alpha$ line of the three observed light-echo
spectra. Both
\specMlo\ and \specDP\ show narrow emission at zero velocity, most
likely caused by the incompletely subtracted fore/background
emission. However, they also show asymmetry at the wavelength for
which we would expect H$\alpha$ emission if the hydrogen has the same
blueshifted velocity of -210~km~s$^{-1}$, indicated with the red
line. If this is indeed a blueshifted H$\alpha$ from the eruption, it
is a narrow one unlike many of the SN impostors, but similar to the
coolest ones like UGC~2773~OT2009-1\cite{Smith10,Foley11}
.

\subsection{Epoch of Spectra}

The slit position of the light echo spectrum is shown in the bottom
right panel of Figure~1.
\ifcheck
  \redpen{CHECK \ref{fig:diffim}} 
\fi
Because the $\eta$~Car Great Eruption lasted over a decade, it is not
straightforward to determine the epochs probed by the light-echo
spectra. In order to attempt this, we compare the historical light
curve\cite{SF11} (see Figure~2)
\ifcheck
  \redpen{CHECK \ref{fig:lc}} 
\fi
with time variation of the light echoes.

We summarize here an account\cite{SF11} of the photometric history of
the Great Eruption of $\eta$~Car.  John Herschel recorded a
brightening of $\sim$1 mag in less than two weeks near the end of
1837. Over the next few months, $\eta$~Car slowly faded
again. Unfortunately, there is a gap in brightness estimates for the
period between late 1838 and 1841.  There was another brightening in
early 1843, and then another episode where $\eta$~Car again faded back
to typical non-outburst brightness. Late in 1844, $\eta$~Car's
greatest recorded brightening episode began, with peak brightness
occurring in early 1845. Thereafter a slow decline began which lasted
for the next decade.

We can compare $\eta$~Car's historical light curve to the light curve
we derived from the light echoes: at a given epoch and position on the
sky, the light-echo flux is the flux of the source event integrated
over a range of epochs. The range of epochs represented in the
instantaneous flux at a given pointing depends on both astrophysical
(dust width and inclination) and observational (seeing) factors, and
can be represented by a near-Gaussian window
function\cite{Rest11_leprofile}. For the light-echo systems we have
characterized around the Cas~A and Tycho remnants, the full width at
half maximum was on the order of 25-120
days\cite{Rest11_leprofile}. Thus if the flux at the same spatial
position is measured at different epochs, the brightness recorded will
be the true outburst light curve convolved with the window function.

With four more epochs of the light echo obtained at the Faulkes
Telescope South (FTS) 2-m telescope, we generated light curves of
$\eta$~Car at 17 positions along the spectroscopic slit shown in
Figure~1.
\ifcheck
  \redpen{CHECK \ref{fig:diffim}} 
\fi
Since we have images from 2003 to mid-2011, these light curves span
$\sim$8 years.  After an overall normalization to each of these light
curves to match each other, the differences in light-curve profiles
are remarkably small, with a typical standard deviation in flux of
2-4\%
(see Table~S5).
\ifcheck
  \redpen{CHECK} 
\fi
We therefore average these light curves into a single light curve. We
show this light curve in Figure~2,
\ifcheck
  \redpen{CHECK \ref{fig:lc}} 
\fi
shifted by 174.2~yrs (green circles), 167.95~yrs (red circles), and
166.28~yrs (blue circles) in order to match the 1838, 1843, and 1845
outbursts, respectively. The earliest epoch (2003) needs to be
considered an upper limit, since that epoch was used as the template
for the difference imaging, and we determined the light-echo flux at
that epoch as the difference between the flux in the template image
and the flux at a different, apparently empty position in the same
image. In other words, there is an undetermined flux zeropoint when an
on-frame reference location may contain light echo flux.
It is also important to note that the light echo light curve is
determined from images in SDSS~$i$, which can introduces systematics
when comparing it to the visual magnitude of the historical light
curve, in particular since LBV's change their colors during outbursts.

The light-echo light curve we derive from our images clearly reveals
brightening of 2 magnitudes or more within an interval of 8 years. The
Lesser Eruption of $\eta$~Car from 1887 to 1896 caused $\eta$~Car to
brighten by roughly 1~magnitude only. Unless the Lesser Eruption was
extremely asymmetric such that it increased by an additional
1~magnitude or more in the direction of the light echo, the Lesser
Eruption is excluded as the source of the light echoes.  We also note
that most of the brightening occurs between the second and third
epochs in a 9 month time-span, indicating that the peak was at some
time between these epochs.

We conclude that there are three remaining eruption epochs to which
our observed light curve could be assigned: 1)~the 1838 peak, 2)~the
1843 peak, and 3)~the 1845 peak. A fourth possibility is an even
earlier, unrecorded brightening. Our ability to decide among the
different epochs relies on the time differences between our images
and spectra. The second- and third-epoch pair have only 9 months
between them and therefore provide strong discrimination for
brightenings. The spectroscopic observations were done shortly after
the third-epoch images, corresponding to $<$9 months after a
brightness maximum.

We now discuss the individual epochs and the likelihood that they
correspond to the brightening observed in our images:
\begin{itemize}
\item {\bf 1838 peak:}
The first two epochs agree very well with the historical light curve.
The historical light curve seems to decline faster after the peak at
the beginning of 1838. Unfortunately, the historical data that
constrain this decline are few and have large uncertainties. In
addition, if the width of the scattering dust filament is very large,
then the window function can span a wide range of epochs, which would
cause the light-echo light curve to have a shallower decline then the
event light curve.
\item {\bf 1843 peak:}
The first two epochs again agree very well. The decline in brightness
is in very good agreement with what would be expected from the
historical light curve.
\item {\bf 1845 peak:}
The first two epochs also agree very well with the 1845 peak.
However, the current decline in brightness ($\approx$0.6 mag
yr$^{-1}$) is considerably faster than the typical decline of
$\approx$0.1mag yr$^{-1}$ observed after 1845. The light-echo light
curve can only be put in agreement with the historical light curve if
there was a fast, unobserved decline from the peak in 1845.
\end{itemize}

We conclude that the most likely source of the light echoes for which
we have spectra is the 1843 outburst, shortly after its peak.
Assuming typical dust properties as observed for light echoes of other
Galactic sources \cite{Rest11_leprofile}, the spectra is the
light-curve-weighted average over an epoch range of 1-4 months.  In
the future, we will be able to further constrain the epoch range with
additional imaging data.  The two peaks in 1838 and 1845 cannot yet be
excluded completely, but continued observations of the light echoes at
monthly intervals should discriminate the possibilities within a year.
This work, combined with targeted new observations and other
contemporary images which may become known, will likely provide a
detailed photometric time-series of the Great Eruption with the
probability of a useful companion spectral time-series.  The recovery
of valuable astrophysical observational data from the pre-imaging era
is another powerful illustration of the power of light-echo
observations

\clearpage


\input{figures}

\input{sifigures}

%% file: figures.tex
\fignature{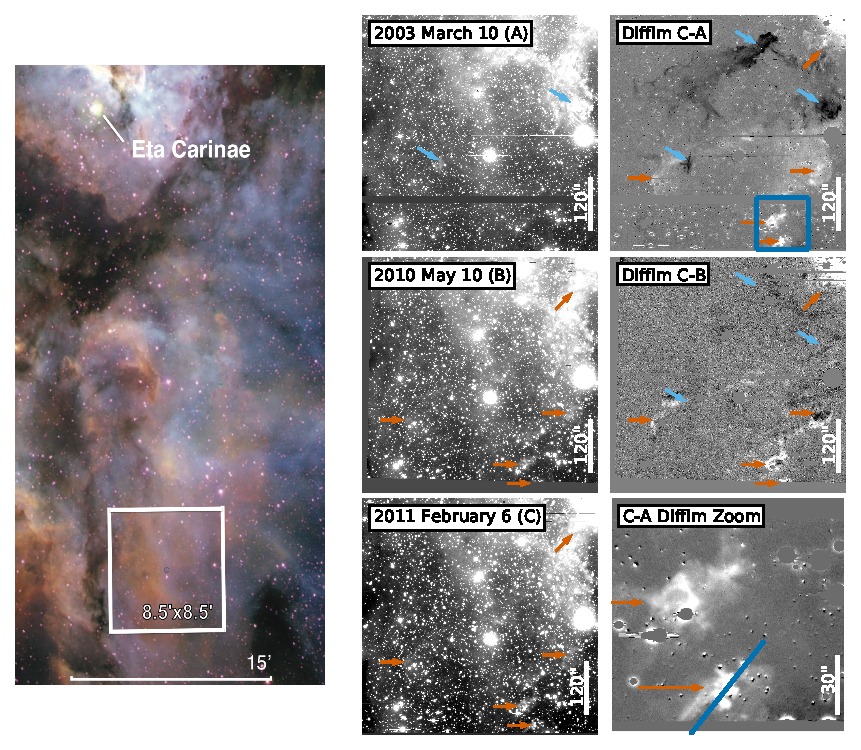}{0.97}
{$\eta$~Car light echoes. The left panel shows the positions of
$\eta$~Car and our images (white box), plotted on an image
in the light of 3 different emission lines:
oxygen (blue), hydrogen (green), and sulfur (red) (credit:
Nathan Smith, University of Arizona/NOAO/AURA/NSF).
%
%
The middle panels show the images obtained with the CTIO 4-m
Blanco telescope of a region $\sim$0.5~degrees to the south of
$\eta$~Car at epochs 2003 March 10 ({\it A}), 2010 May 10 ({\it B}),
and 2011 Feb 6 ({\it C}), from top to bottom.  The right panels show
the difference images {\it C-A} and {\it C-B} at the top and middle,
respectively. Example light-echo positions are indicated with blue
(epoch {\it A}) and red (epoch {\it B} and {\it C}) arrows.  The
bottom right panel shows a zoom of the spectrograph slit, indicated
with a blue line.  For all panels north is up and east is to the left.
Applying the vector method that previously allowed us to identify the
source of the light echoes from the supernovae (SNe) that produced the
SN remnants 0509$-$67.5, Cas~A, and Tycho\cite{Rest05b,Rest08b}, we
find that a dramatic brightening of $\eta$~Car must be the origin.  In
these echoes, unlike those of Galactic SNe\cite{Rest11_leprofile},
there is still significant spatial overlap even at separations of one
light-year, suggesting that the duration of the event causing them
must be significantly longer than one year.  We also see brightening
of 2 magnitudes or more within 8 years.  Thus, the Lesser Eruption
from 1887 to 1896, which brightened by only a magnitude, is excluded
as the source.
}
{fig:diffim}

\fignature{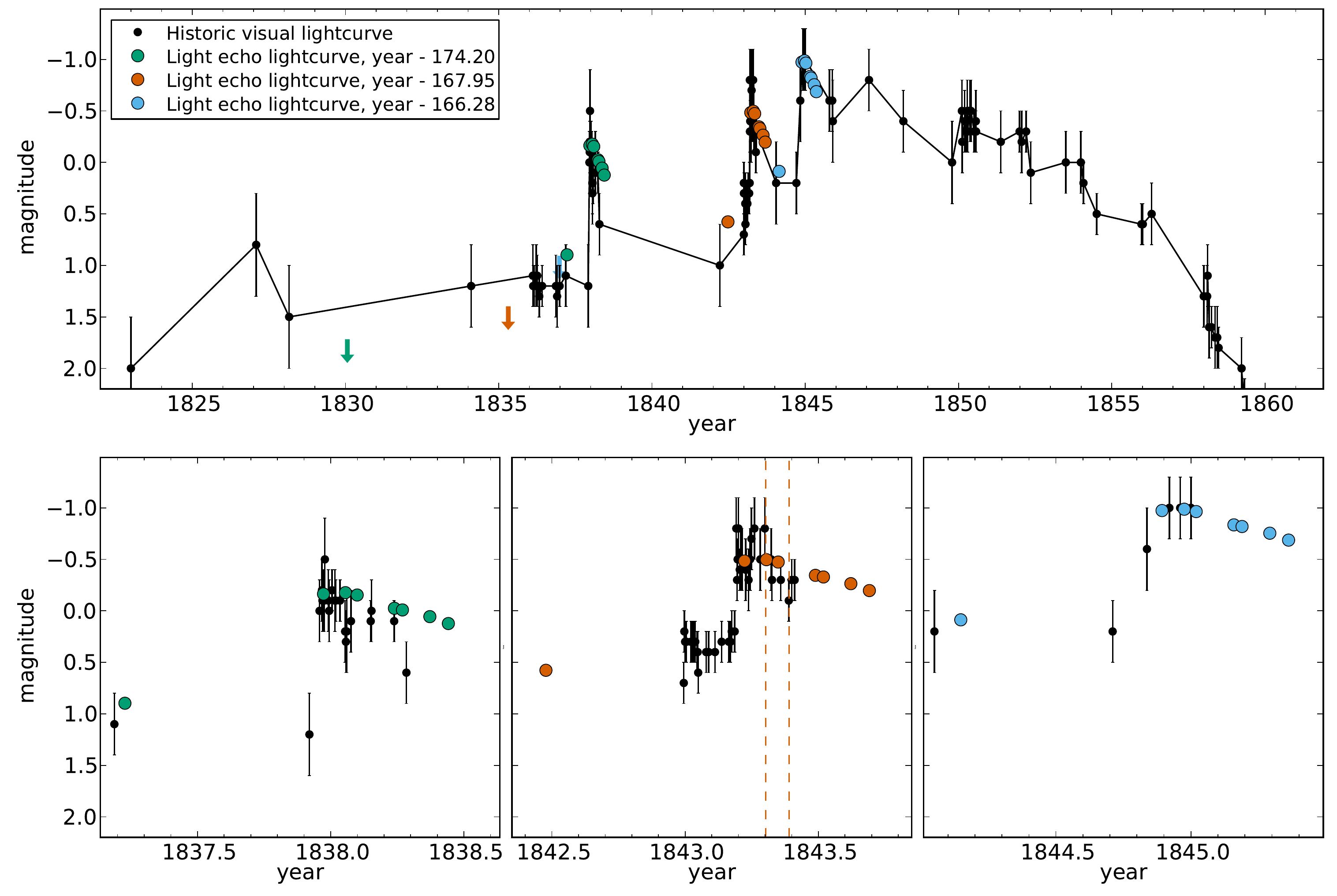}{0.53}
{Historical and light-echo lightcurve of $\eta$~Car. The historical
light curve\cite{SF11} in visual apparent magnitudes is shown with
black circles and lines, with error bars indicating approximate
uncertainties in these eye estimates.  Light echo brightnesses (SDSS
$i$; error bars are the s.d.) from our eight modern images spanning
$\sim$8~years are displayed shifted by 174.2~yrs (green circles),
167.95~yrs (red circles), and 166.28~yrs (blue circles), in order to
illustrate the best-matching time delays for the 1838, 1843, and 1845
outbursts, respectively.
The first epoch is an upper limit indicated with an arrow. 
The upper panel shows the full time range of the Great
Eruption and therefore shows all three potential matches, whereas the
lower panels show the brightnesses from seven of our eight modern
epochs in a magnified time period around each peak.
}
{fig:lc}

\fignature{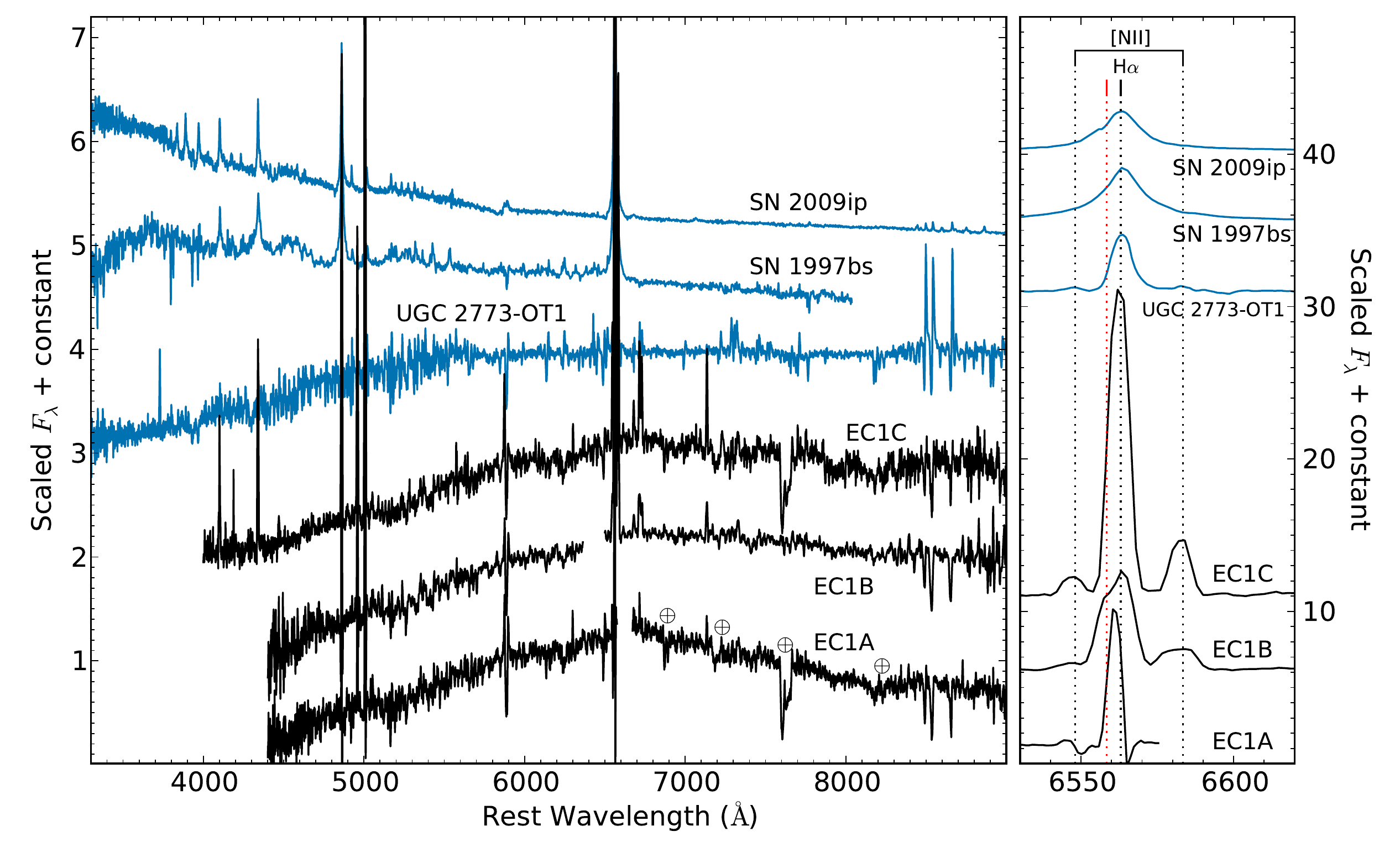}{0.63}
{Light echo spectra of $\eta$~Car's Great Eruption.  The three optical
low-resolution spectra of the light echo (black lines) were taken at
J2000 position RA=10:44:12.127 and Decl.=$-$60:16:01.69 in March and
April 2011 obtained at the Magellan~I 6.5-m and du~Pont 2.5-m
telescopes of the Las Campanas Observatory, Chile. 
A log of the spectroscopic observations and
details of the spectra is presented in Table~S1.
\ifcheck
  \redpen{CHECK} 
\fi
The slit positions differ only slightly in slit angle.
The spectra were reduced using standard techniques and then
wavelength-calibrated using observations of an HeNeAr lamp. The
wavelength calibration was checked and corrected using night-sky
emission lines, especially \ion{O}{I} 5577\AA, and OH lines in the red
part of the spectrum. We flux-calibrated the spectra using a flux
standard observed the same night as the science observations.  The
left panel shows the spectra from 5000 to 9000\AA.  The spectra are
not corrected for reddening nor for the blue-ward scattering by the
dust.  The blue lines show for comparison spectra of three examples of
SN imposters: SN~1997bs, SN~2009ip, and UGC~2773-OT1.  The right panel
shows the H$\alpha$ and [\ion{N}{II}] emission lines. Note that the
background emission-line subtraction is incomplete since they
spatially vary. Also, for
\specMhi\ H$\alpha$ is at the edge of the chip and is therefore uncertain.
}
{fig:spec}

\fignature{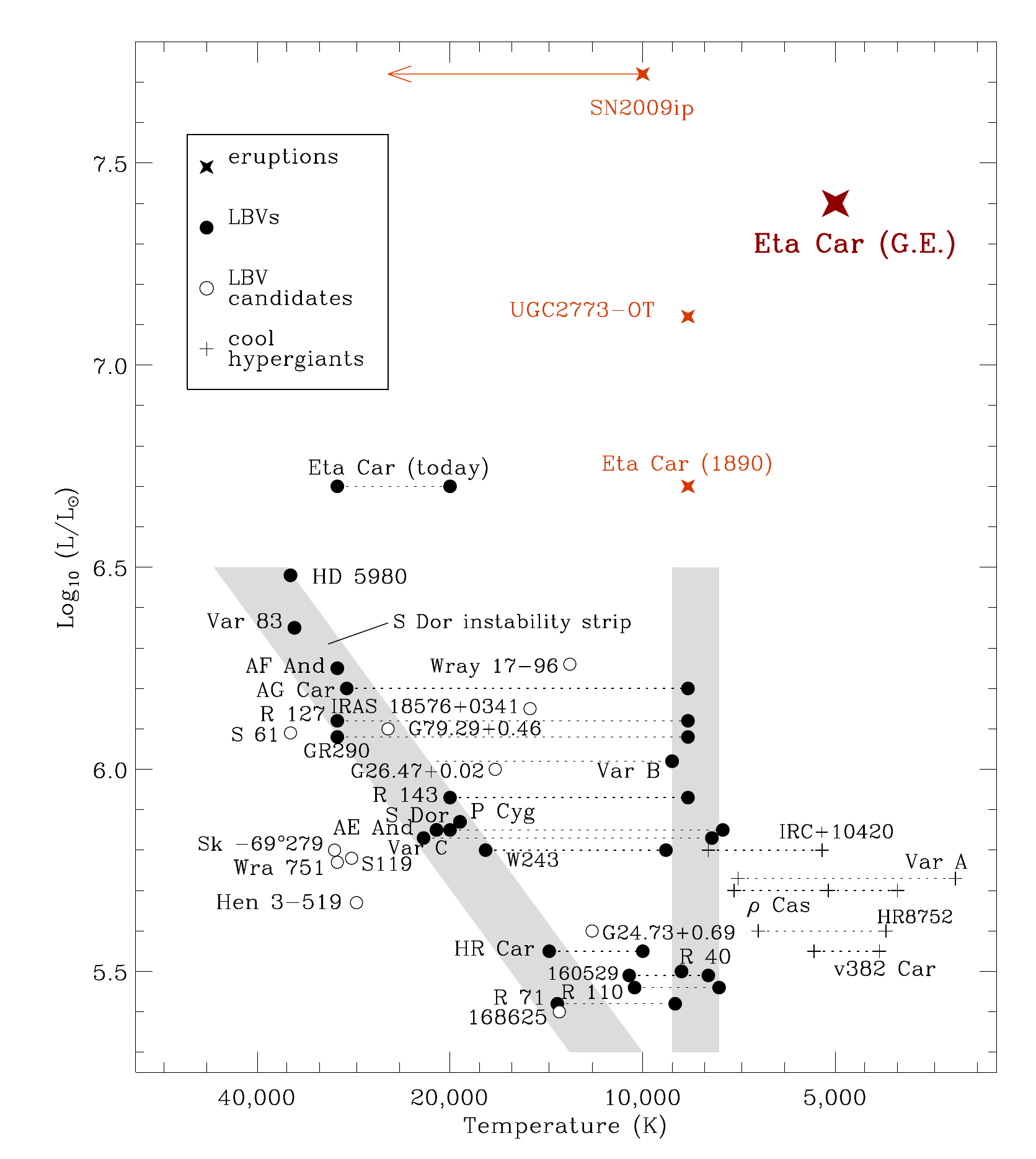}{0.8}
{HR diagram with LBVs and $\eta$~Car.
Adaptation\cite{Smith04} of an HR diagram showing LBVs, related
hypergiant stars, and the peak luminosities of LBV-like eruptions. The
grey bands denote the typical locations of LBVs in quiescence and in S
Doradus excursions.  Temperatures for the Great Eruption and 1890
eruption of $\eta$~Car are based on the echo spectra presented here
and the F-type spectrum of the 1890 event\cite{Walborn77},
respectively.
The temperature of 10,000~K for SN~2009ip is based on the observed
continuum shape, but this is only a lower limit because of the
possible effects of circumstellar or host galaxy
reddening\cite{Smith10}.  Because of the presence of \ion{He}{I} lines
in the spectrum, the true temperature is probably much hotter.
The 8500~K temperature of UGC2773-OT is indicated by the F-type
absorption features in its spectrum, and this temperature is
relatively independent of reddening\cite{Smith10,Foley11}.
}
{fig:hr}

%% file: sifigures.tex
\ifastroph
   \setcounter{figure}{0}
   \renewcommand{\thefigure}{S\arabic{figure}}
\else
  \ifsuppl
     \setcounter{figure}{-4}
  \fi
\fi

\fignature{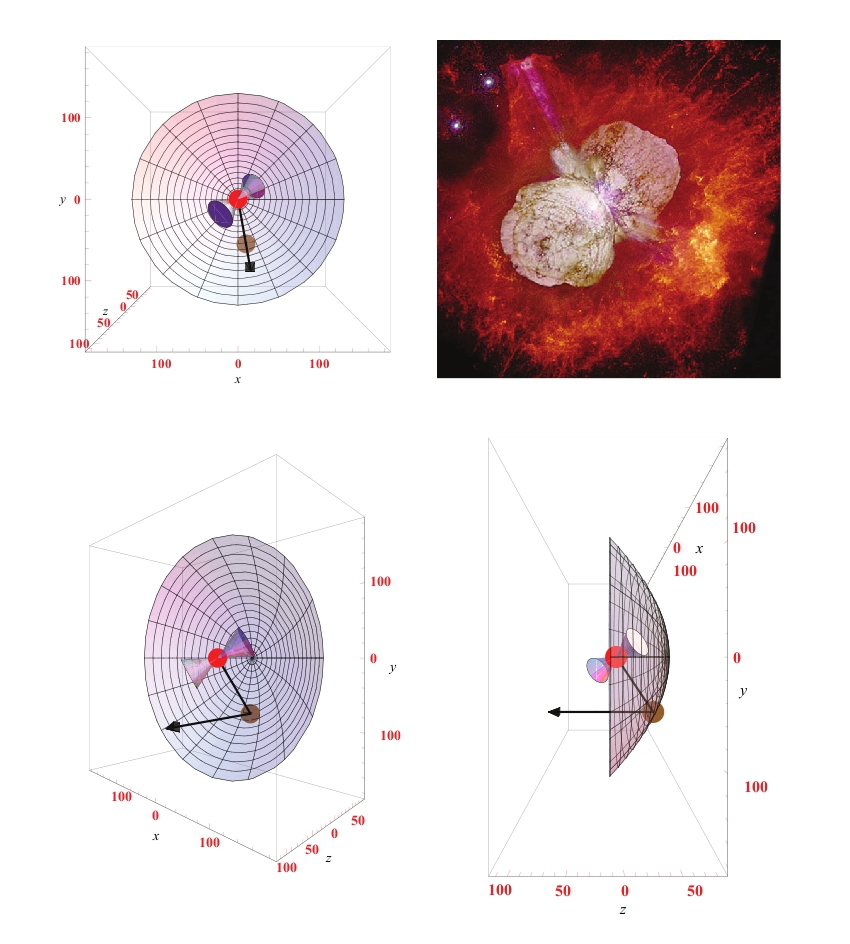}{0.98}
{3D-plot of the light path. North is toward
the positive $y$-axis (up), east is toward the negative $x$-axis
(left), and the positive $z$-axis points toward the observer with the
origin at $\eta$~Car.  The red and brown circles indicate $\eta$~Car
and scattering dust, respectively.  The parabolic relation between the
spatial parameters of the scattering dust and the time since outburst
is described by the well-known light echo equation\cite{Couderc39}:
Assuming a time since outburst of 169 years, and a distance of 7660
light-years\cite{Smith06}, we find that the scattering dust is at a
position ($x$,$y$,$z$)=(14,-78,-66) in light-years.  
The black lines show the path of
the light scattering from the light-echo-producing dust
concentrations.  The top-right panel shows an \textit{HST\/} image
(credit: Nathan Smith / Jon Morse / NASA) of $\eta$~Car, which is
surrounded by expanding lobes of gas denoted as the Homunculus Nebula
visible to the NW and SE. These lobes were created by the 1838-1858
Great Eruption. We indicate the lobes in the 3D plots with the two
cones (Note that the size of the cones is not to scale).
}
{fig:3Dplot}

\fignature{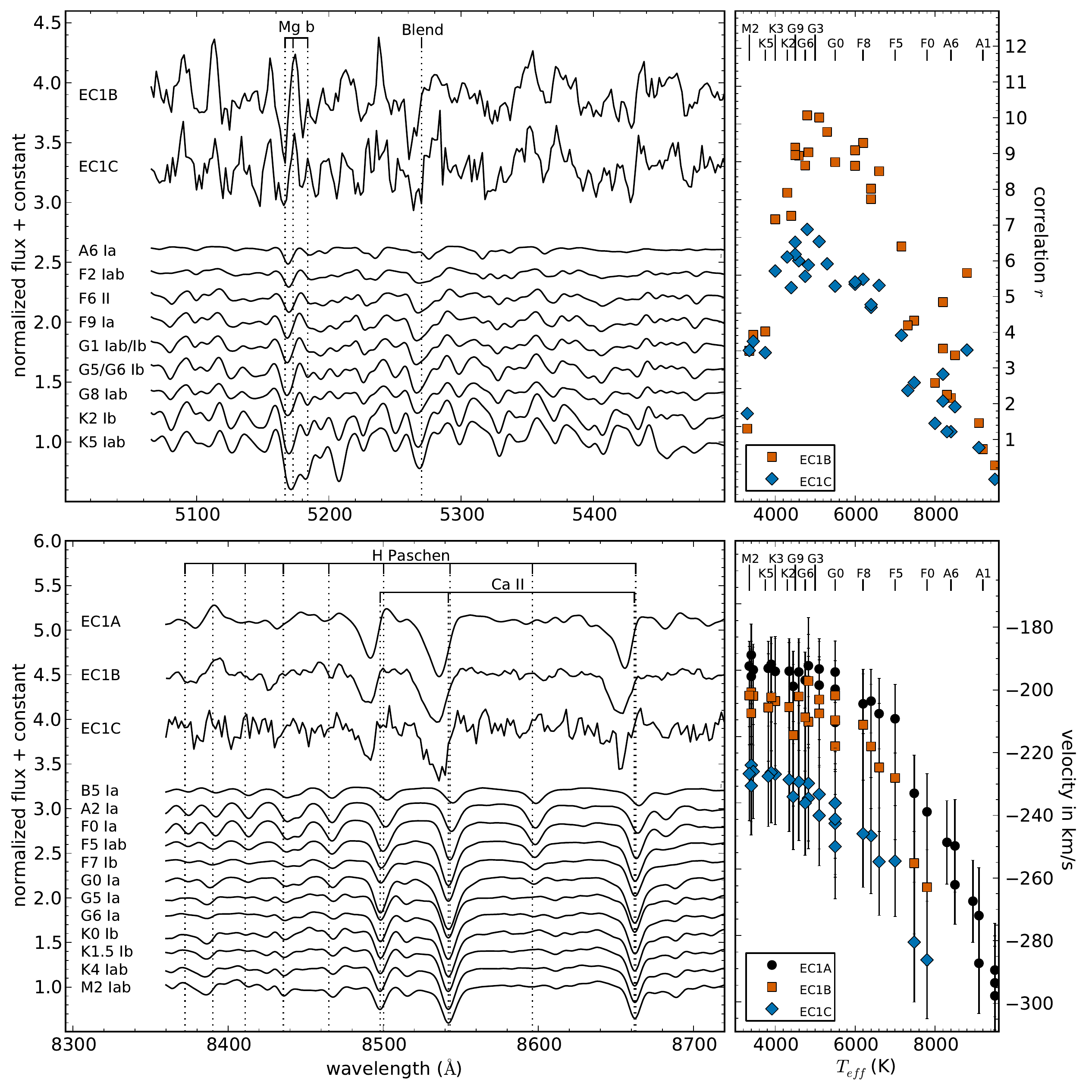}{0.57}
{  Spectral
  type comparison. The stellar comparison spectra are convolved with a
  Gaussian of FWHM 7~\AA\ so that their resolution matches those 
  of \specMlo\ and \specDP. Similarly, we convolve
  the spectrum \specMhi, which has an original resolution of 4\AA,
  with a Gaussian of FWHM 5.7~\AA. We then divide the spectra by
  the low-order continuum which we determine by convolving the spectra
  with a Gaussian of FWHM 200~\AA. The upper left panel
  compares two of the observed light echo
  spectra to a selection of UVES supergiants\cite{Bagnulo03} in the 5060-5500~\AA\
  wavelength range (We do not show spectrum \specMhi\ since its S/N
  ratio in this blue wavelength range is too low).  The light echo
  spectra correlate very well with late-F and G-type stars, in
  particular the Mg~b lines, and the \ion{Ca}{I}, \ion{Fe}{I},
  \ion{Ti}{I}, \ion{Cr}{I} blend at 5270~\AA. The upper right panel
  shows the cross-correlation between the light echo and the UVES
  spectra in the wavelength range of 5050-6500~\AA\  
  as calculated by
  the IRAF routine {\tt xcsao}.
%
  Since the UVES spectra have a gap around the \ion{Ca}{II}
  IR triplet, we use the \ion{Ca}{II} triplet spectral
  library\cite{Cenarro01} in that wavelength range, as shown in the
  bottom left panel.
  The bottom right panel shows the blue-shifted velocity determined by {\tt
  xcsao} from the \ion{Ca}{II} triplet with respect to the effective temperature\cite{Humphreys84}
  of the supergiant template spectra. 
}
{fig:speclines}

\fignature{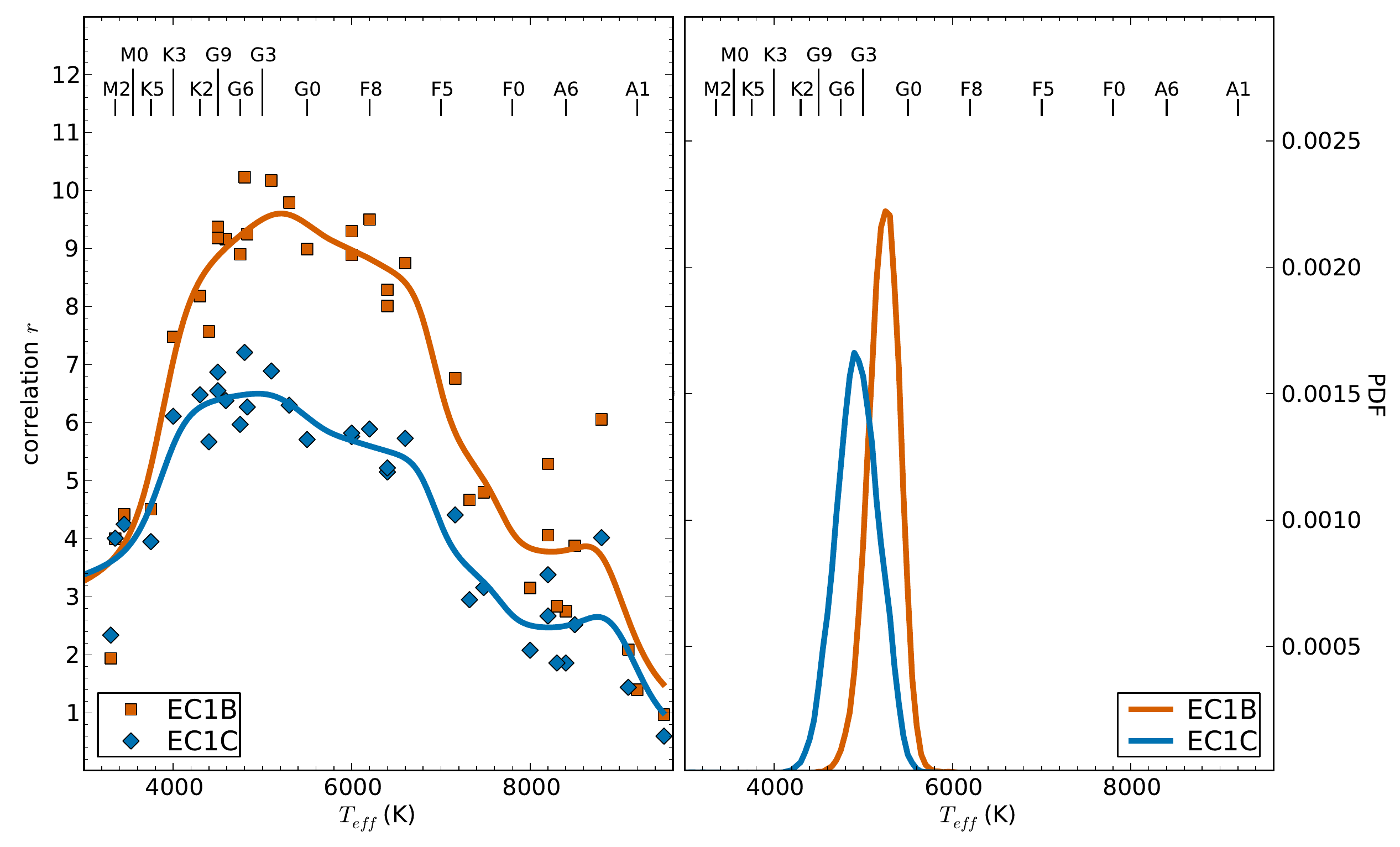}{0.63}
{Effective temperature of the light echo spectrum.  The left panel
shows the cross-correlation between the light echo and the UVES
spectra as calculated by the IRAF routine {\tt xcsao} for \specMlo\
(red squares) and \specDP\ (blue diamonds).  The lines indicate $r(T)$
smoothed with a Gaussian of width $\sigma=300$~K, peaking at 5210~K
and 4950~K for \specMlo\ (red) and \specDP\ (blue), respectively. The
right panel shows the PDF of the best-correlating supergiant
temperature, determined by bootstrap resampling the $r$ distribution
$10^5$~times. The 95\% temperature confidence intervals are
4850-5550~K and 4450-5400~K for \specMlo\ and \specDP, respectively.}
{fig:pdf}

%% file: tables.tex
\begin{sidewaystable}
\footnotesize
\begin{center}
\begin{tabular}{cclcccccccl}
\hline
Name & 
UT Date & 
Telescope & 
Instrument & 
Exptime &
Grism & 
Slitwidth &
Range & 
FWHM resolution & 
P. A. & 
Observer \\
 & 
 & 
 & 
 & 
(sec) &
(lines/mm) & 
($\arcsec$) &
(\AA) & 
(\AA) & 
($\deg$) & 
\\
\hline
EC1A & 2011/04/06    &  Magellan~I 6.5m & IMACS & $2\times 1800$  & 300  & 0.9 & $3800-10000$ & 4 & 339 & J. Prieto\\
EC1B & 2011/03/07    &  Magellan~I 6.5m & IMACS & $2\times 1800$  & 200  & 0.9 & $4000-10000$ & 8 & 324 & R. Chornock\\
&   &   &  &  &   & &  & &  & \& R. Foley\\
&   &   &  &  &   & &  & &  & \& W. Fong\\
EC1C & 2011/04/08    &  du~Pont 2.5m    & WFCCD & $5\times 1800$  & 400  & 1.7 & $3700-9200$  & 7 & 339 & J. Prieto \\
\hline\label{tab:spec}
\end{tabular}
\caption{Supplementary Information: Log of Spectroscopic Observations }
\end{center}
\end{sidewaystable}

\begin{table}
\begin{center}
\begin{tabular}{lllc}
\hline
Star Name & HD & Spectral Type & $T_{\rm eff}$\\
\hline
\hline
N Car          & HD 47306    &    A0      II &         9500\\
HR 4541        & HD 102878   &    A2     Iab &         9100\\
$\dots$        & HD 34295    &    A4      II &         8800\\
n Vel          & HD 74272    &    A5      II &         8500\\
y Car          &  HD 97534   &    A6      Ia &         8400\\
HR 3739        & HD 81471    &    A7     Iab &         8300\\
$\iota$ Car    & HD 80404    &    A8      Ib &         8200\\
$\dots$        & HD 104111   &    A8      II &         8200\\
V399 Car       & HD 90772    &    A9      Ia &         8000\\
HR 3496        &  HD 75276   &    F2     Iab &         7480\\
b Vel          &  HD 74180   &    F3      Ia &         7320\\
HR 5024        & HD 115778   &    F4      II &         7160\\
$\rho$ Pup     & HD 67523    &    F6      II &         6600\\
BG Cru         & HD 108968   &    F7   Ib/II &         6400\\
HR 8470        & HD 210848   &    F7      II &         6400\\
$\delta$ CMa   &  HD 54605   &    F8     Iab &         6200\\
V810 Cen       & HD 101947   &    F9     Iab &         5900\\
$\gamma^1$ Nor & HD 146143   &    F9      Ia &         5900\\
BB Sgr         & HD 174383   &    G0      Ib &         5500\\
ER Car         &  HD 97082   &    G1  Iab/Ib &         5300\\
$\dots$        & HD 136537   &    G2      II &         5100\\
$\beta$ Crv    & HD 109379   &    G5      II &         4830\\
$\dots$        & HD 125809   & G5/G6      Ib &         4790\\
HR 3673        &  HD 79698   &    G6      II &         4750\\
$\tau$ Leo     &  HD 99648   &    G8     Iab &         4590\\
d Cen          & HD 117440   &    G9      Ib &         4500\\
HR 3583        &  HD 77020   &    G9      II &         4500\\
HR 554         &  HD 11643   &    K1      II &         4400\\
$\epsilon$ Peg & HD 206778   &    K2      Ib &         4300\\
3 Cet          & HD 225212   &    K3     Iab &         4000\\
HR 611         &  HD 12642   &    K5     Iab &         3750\\
HR 2508        & HD 49331    &    M1      Iab &        3450\\
V528 Car       & HD 95950    &    M2      Ib &         3350\\
CR Cir         & HD 131217   & M2/M3      II &         3300\\

\hline\label{tab:UVESstellarlib}
\end{tabular}
\caption{Supplementary Information: UVES Supergiant Stellar Spectra Library\cite{Bagnulo03}. We
convert the spectral type into the effective temperature $T_{\rm eff}$ 
using the
spectral-type to temperature relation for
supergiants \cite{Humphreys84}}
\end{center}
\end{table}

\begin{table}
\begin{center}
\begin{tabular}{lllcc}
\hline
Star Name & HD & Spectral Type &  $T_0$ &  $T_{\rm eff}$\\
\hline
\hline
5 Per         & HD 13267   &   B5     Ia &	 13800 &	13700\\
$\nu$ Cep     & HD 207260  &   A2     Ia &	  9100 &	 9100\\
$\phi$ Cas    & HD 7927    &   F0     Ia &	  7425 &	 7800\\
$\nu$ Aql     & HD 182835  &   F2     Ib &	  7350 &	 7480\\
44 Cyg        & HD 195593  &   F5    Iab &	  6600 &	 7000\\
35 Cyg        & HD 193370  &   F6     Ib &	  6200 &	 6600\\
V440 Per      & HD 14662   &   F7     Ib &	  5900 &	 6400\\
HR 7008       & HD 172365  &   F8  Ib-II &	  5500 &         6200\\
$\dots$       & HD 18391   &   G0     Ia &	  5500 &	 5500\\
HR 7456       & HD 185018  &   G0     Ib &	  5550 &	 5500\\
14 Per        & HD 16901   &   G0  Ib-II &	  5478 &	 5500\\
HR 3459       & HD 74395   &   G2    Iab &	  5250 &	 5100\\
$\beta$ Dra   & HD 159181  &   G2    Iab &	  5250 &	 5100\\
$\dots$       & HD 187299  &   G5     Ia &	  5010 &	 4830\\
$\beta$ Lep   & HD 36079   &   G5     II &	  5170 &	 4830\\
$\xi$ Pup     & HD 63700   &   G6     Ia &	  4990 &	 4750\\
$\xi^1$ Cet   & HD 13611   &   G8    Iab &	  5040 &	 4590\\
$\dots$       & HD 12014   &   K0     Ib &	  5173 &	 4450\\
$\zeta$ Cep   & HD 210745  & K1.5     Ib &	  4500 &	 4350\\
$\gamma^1$ And& HD 12533   &   K3    IIb &	  4383 &	 4000\\
41 Gem        & HD 52005   &   K4    Iab &	  4116 &	 3900\\
V809 Cas      & HD 219978  & K4.5     Ib &	  4250 &	 3825\\
$\alpha$ Ori  & HD 39801   &   M2    Iab &	  3614 &	 3350\\
\hline\label{tab:CaIIstellarlib}
\end{tabular}
\caption{Supplementary Information: \ion{Ca}{II} 
IR triplet Supergiant Stellar Spectra
Library\cite{Cenarro01}.  The temperature $T_0$ is the effective
temperature quoted by Cenarro et al.\cite{Cenarro01}.  For consistency
we also calculate $T_{\rm eff}$ using the same stellar type-temperature
relation\cite{Humphreys84} we used for the UVES spectral library.
}
\end{center}
\end{table}

\begin{table}
\begin{center}
\begin{tabular}{lcc}
\hline
Line & $\lambda_{min}$ & $\lambda_{max}$ \\
\hline
\hline
\ion{O}{I} 5577.4 & 5572 & 5582 \\
\ion{Na}{I}~D 5890.0, 5895.9 & 5860 & 5921 \\
\ion{O}{I} 6300.2 & 6295 & 6305  \\
\ion{O}{I} 6363.9 & 6359 & 6369 \\
\hline\label{tab:excludedlines}
\end{tabular}
\caption{
List of the fore/background emission lines which are
excluded from the 5050-6500~\AA wavelength range of the correlation
analysis. For each spectroscopic line, $\lambda_{min}$ and
$\lambda_{max}$ define the initial and final wavelengths in \AA~of the
contaminating line region we exclude.
}
\end{center}
\end{table}

\begin{sidewaystable}
\scriptsize
\begin{center}
\begin{tabular}{lccccccccc}
\hline
& 03/10/2003 & 05/10/2010 & 02/06/2011 & 03/08/2011 & 03/24/2011 & 05/14/2011 & 05/25/2011 & 07/01/2011 & 07/27/2011\\
\hline
\hline
Instrument & CTIO4m Mosaic II & CTIO4m Mosaic II & CTIO4m Mosaic II & Magellan IMACS f2 & FTS & FTS & FTS & CTIO4m Mosaic II & FTS\\
MJD & 52708.11960 & 55326.01102 & 55598.22252 & 55628.26319 & 55644.39484 & 55695.38567 & 55706.41416 & 55743.98288 & 55769.35728\\
Exp. time & 60 & 120 & 120 & 30 & 600 & 1200 & 600 & 120 & 600\\
F0 & $1.40\pm0.13$ & $\dots$ & $6.86\pm0.19$ & $6.82\pm0.21$ & $6.95\pm0.21$ & $6.22\pm0.18$ & $6.15\pm0.19$ & $5.66\pm0.18$ & $5.67\pm0.21$\\
F1 & $1.02\pm0.11$ & $\dots$ & $7.45\pm0.17$ & $7.20\pm0.18$ & $6.85\pm0.18$ & $6.16\pm0.16$ & $6.13\pm0.17$ & $5.45\pm0.15$ & $5.10\pm0.18$\\
F2 & $1.27\pm0.11$ & $\dots$ & $7.39\pm0.16$ & $6.92\pm0.17$ & $6.88\pm0.17$ & $6.14\pm0.15$ & $6.10\pm0.16$ & $5.62\pm0.15$ & $5.26\pm0.17$\\
F3 & $1.32\pm0.11$ & $\dots$ & $7.32\pm0.16$ & $6.86\pm0.18$ & $6.86\pm0.18$ & $6.04\pm0.15$ & $6.09\pm0.16$ & $5.76\pm0.15$ & $5.42\pm0.18$\\
F4 & $1.22\pm0.11$ & $\dots$ & $7.27\pm0.16$ & $7.00\pm0.18$ & $6.93\pm0.18$ & $6.16\pm0.15$ & $6.08\pm0.16$ & $5.58\pm0.15$ & $5.37\pm0.18$\\
F5 & $1.39\pm0.10$ & $\dots$ & $7.26\pm0.15$ & $6.93\pm0.16$ & $6.84\pm0.16$ & $6.10\pm0.14$ & $6.12\pm0.15$ & $5.58\pm0.14$ & $5.54\pm0.16$\\
F6 & $1.38\pm0.09$ & $\dots$ & $7.31\pm0.13$ & $7.31\pm0.13$ & $7.06\pm0.14$ & $6.06\pm0.12$ & $5.97\pm0.12$ & $5.57\pm0.12$ & $5.17\pm0.13$\\
F7 & $1.36\pm0.08$ & $\dots$ & $7.15\pm0.12$ & $6.98\pm0.13$ & $6.80\pm0.13$ & $6.13\pm0.11$ & $6.09\pm0.12$ & $5.85\pm0.11$ & $5.34\pm0.13$\\
F8 & $1.06\pm0.09$ & $\dots$ & $6.66\pm0.13$ & $6.64\pm0.14$ & $6.95\pm0.14$ & $6.21\pm0.12$ & $6.14\pm0.13$ & $5.90\pm0.12$ & $5.82\pm0.14$\\
F9 & $1.07\pm0.09$ & $\dots$ & $6.61\pm0.13$ & $6.87\pm0.14$ & $6.94\pm0.14$ & $6.25\pm0.12$ & $6.14\pm0.13$ & $5.93\pm0.12$ & $5.60\pm0.14$\\
F10 & $1.13\pm0.08$ & $\dots$ & $6.58\pm0.12$ & $7.00\pm0.13$ & $7.04\pm0.13$ & $6.30\pm0.11$ & $6.25\pm0.12$ & $5.78\pm0.11$ & $5.41\pm0.13$\\
F11 & $1.13\pm0.08$ & $\dots$ & $6.98\pm0.11$ & $7.32\pm0.13$ & $6.98\pm0.12$ & $6.28\pm0.11$ & $6.00\pm0.11$ & $5.67\pm0.11$ & $5.23\pm0.12$\\
F12 & $1.27\pm0.08$ & $2.61\pm0.16$ & $6.95\pm0.11$ & $7.56\pm0.12$ & $7.04\pm0.12$ & $6.15\pm0.11$ & $5.92\pm0.11$ & $5.68\pm0.11$ & $5.21\pm0.12$\\
F13 & $1.43\pm0.09$ & $2.65\pm0.12$ & $6.93\pm0.12$ & $7.40\pm0.14$ & $6.90\pm0.13$ & $6.05\pm0.12$ & $6.02\pm0.12$ & $5.83\pm0.11$ & $5.32\pm0.13$\\
$\bar{F}$ & 1.24 & 2.64 & 7.01 & 7.09 & 6.94 & 6.17 & 6.08 & 5.72 & 5.38\\
$\sigma_{\bar{F}}$ & 0.14 & 0.02 & 0.29 & 0.25 & 0.08 & 0.08 & 0.08 & 0.14 & 0.20\\
\hline\label{tab:lc}
\end{tabular}
\caption{
The individual lightcurves are normalized so that the average $-2.5
\log_{10}(F) = -2.0$ for MJD in 55590-555660.  For the image from
05/10/2010, 12 out of 14 positions fell into the chip gap and could
not be determined.
}
\end{center}
\end{sidewaystable}